\documentclass[a4paper,12pt]{article}
\usepackage{amsmath,graphicx}       
\renewcommand{\L}{\Lambda}

\begin{document}
\begin{titlepage}
\begin{flushright}
IFUM 621/FT\\
May 1998\\
\end{flushright}
\vspace{1.5cm}
\begin{center}
{\bf \large HARD-SOFT RENORMALIZATION OF THE MASSLESS WESS-ZUMINO MODEL}
\footnote{Work supported in part by M.U.R.S.T. and INFN. }\\
\vspace{1 cm}
{ M. PERNICI} \\ 
\vspace{2mm}
{\em INFN, Sezione di Milano, Via Celoria 16, I-20133 Milano, Italy}\\
\vspace{0.6 cm}
{ M. RACITI and F. RIVA}\\
\vspace{2mm}
{\em Dipartimento di Fisica dell'Universit\`a di Milano\\
and\\ INFN, Sezione di Milano,\\ Via Celoria 16, I-20133 Milano, Italy}\\
\vspace{2cm}
\bf{ABSTRACT}
\end{center}
\begin{quote}
We show that in a Wilsonian renormalization scheme with zero-momentum
subtraction point the massless Wess-Zumino model satisfies the
non-renormalization theorem; the finite renormalization of the
superpotential appearing in the usual non-zero momentum subtraction
schemes is thus avoided.

We give an exact expression of the beta and gamma functions in terms of the
Wilsonian effective action; we prove the expected relation 
$\beta = 3g\gamma$.
 
We compute the beta function at the first two loops,
finding agreement with previous results.
\end{quote}
\end{titlepage}
\section*{}

The perturbative renormalization of massless theories presents
difficulties of a practical order, due to the presence of infrared 
singularities.

In theories with symmetries admitting an invariant regulator these
difficulties can be avoided choosing an appropriate subtraction
scheme, like the minimal subtraction scheme \cite{tH}.

In some theories, however, these procedures cannot be adopted and the
Ward identities must be verified explicitly in presence of massless
fields. This is for instance the case of chiral gauge theories.

A possible way of renormalizing these theories is to impose
renormalization conditions at non-zero momentum subtraction points  \cite{Rm}; however
this scheme is computationally awkward, since it is usually technically hard
to satisfy the Ward identities at non-zero momenta, especially the
Slavnov-Taylor identities.

In the hard-soft renormalization schemes, first introduced in \cite{LM} 
with the purpose of studying in a simple way the renormalizability of
massless theories with BPHZ, a splitting of the fields into hard and
soft fields is made at a scale $\L_R$, in such a way that the
renormalization conditions can be chosen at zero momentum.

A recent discussion of the hard-soft (HS) schemes in the Wilsonian
approach \cite{Wil,Polchi} can be found in \cite{PRR,PR}.

In massless QED the renormalization conditions can be chosen at zero
momentum at a Wilsonian renormalization scale $\L_R$ \cite{PRR}, satisfying
effective gauge and axial Ward identities; the usual Ward identities follow
automatically for any ultraviolet cut-off.
 The nice feature of this approach is
that the effective Ward identities are easily satisfied, being the
renormalization conditions imposed at zero momentum.
Detailed one-loop computations are made in this scheme.
In \cite{Ell} one-loop computations are made in a HS scheme in
Yang-Mills, using dimensional regularization; in this way the gauge Ward
identities are trivially satisfied.

The renormalization group equation can be easily obtained in the HS
schemes, expressing the beta and gamma functions in terms of the
Wilsonian effective action at $\L_R$; in \cite{PR} this
procedure has been applied to the case  of massless $g\phi^4$ and of
massive $g\phi^4$, renormalized in a mass-independent way.

In this letter we apply the HS renormalization scheme to the massless
Wess-Zumino model \cite{WZ}. In this case the supersymmmetric Ward identities
and the R-symmetry are trivially satisfied, choosing an ultraviolet
momentum cut-off.

An interesting feature of the massive Wess-Zumino model is that,
choosing zero-momentum renormalization conditions, the superpotential
is not renormalized \cite{IZ}; as a consequence the simple relation 
$\beta = 3 g \gamma_{\Phi}$ \cite{IWZ} between the renormalization group
functions holds.
Using supergraph Feynman rules, these facts follow from the
non-renormalization theorem \cite{GSR}, stating that all 1PI
graphs  are of the form of a single
integral in superspace; as a consequence all  $1$PI graphs
contributing to the superpotential of the massive Wess-Zumino model
vanish at zero momentum.
In the massless case zero-momentum renormalization conditions
cannot be chosen due to infrared singularities; using non-zero subtraction points
on the two- and three- point Green functions one gets a finite
renormalization of the superpotential \cite{Piguet}.
In fact, starting from two loops, the three-point function is
non-vanishing at generic momenta \cite{TvN,JJW}.

We show that in the HS scheme the non-renormalization property of the
superpotential of the massless Wess-Zumino model is easily satisfied,
and the relation $\beta = 3 g \gamma_{\Phi}$ holds exactly.
We compute the two-loop beta function in the HS scheme, finding
agreement with \cite{AG}, where minimal subtraction was used.
\section*{}
Consider the massless $g\Phi^3$ Wess-Zumino model \cite{WZ} in Euclidean four
dimensional space; the path-integral is
\begin{equation}
Z_{0\L_0}\left[J,\bar{J}\right]=\int {\cal D}\Phi {\cal D}\bar{\Phi}~
\exp\ \left\{-S\left[\Phi,\bar{\Phi}\right] + \int d^6 z J \Phi +
\int d^6 \bar{z} \bar{J} \bar{\Phi}\right\}
\end{equation}
with the bare action
\begin{equation}\label{bare}
S\left[\Phi,\bar{\Phi}\right] = -\int d^8 z \bar{\Phi}K_{\L_0}^{-1}\Phi
+S^I\left[\Phi,\bar{\Phi}\right] 
\end{equation}
\begin{equation}\label{ba}
S^I\left[\Phi,\bar{\Phi}\right] = \int d^8 z c_1 \bar{\Phi}\Phi
+\int d^6 z \frac{c_2}{3!}\Phi^3 +
\int d^6 \bar{z} \frac{c_2}{3!}\bar{\Phi}^3
\end{equation}
We use the superspace conventions of \cite{GGRS}.
In the loop expansion,
at tree level the bare coefficients are $c_1^{(0)}=0$, $c_2^{(0)}=g$.
In momentum space
the cut-off function $K_{\L}(p) = K(\frac{p^2}{\L^2})$ satisfies $K(0)=1$ and goes to
zero at least as fast as $1/x$ for 
$x \to \infty$. 
(In \cite{PRR,PR} $K(x)$ was required to go to zero at least as fast as
$1/x^2$ for $x \to \infty$; relying on the cancellation of quadratic
divergences in supersymmetry, it is sufficient to impose the former
weaker condition). 
$\L_0$ is the ultraviolet cut-off.

This cut-off is compatible with supersymmetry and the R-symmetry, 
so that the counterterms in (\ref{ba}) are the only ones allowed by these
symmetries.

Let us consider the path-integration on the hard modes
\begin{equation}\label{ZL}
Z_{\L\L_0}\left[J,\bar{J};\chi\right]=\int {\cal D}\Phi {\cal D}\bar{\Phi}~
\exp \left\{-S_{\L,\L_0}\left[\Phi,\bar{\Phi};\chi\right] + \int d^6 z J \Phi +
\int d^6 \bar{z} \bar{J} \bar{\Phi}\right\}
\end{equation}
with action
\begin{equation}\label{Lbare}
S_{\L,\L_0}\left[\Phi,\bar{\Phi};\chi\right] =
-\int d^8 z \bar{\Phi}K_{\L\L_0}^{-1}\Phi + S^I\left[\Phi,\bar{\Phi}\right]
-\chi \int d^8 z \bar{\Phi}K_{\L}K_{\L\L_0}^{-1}\Phi
\end{equation}
where the term with real parameter $\chi$ has been introduced for later 
convenience; 
$Z_{\L\L_0}\left[J,\bar{J}\right] \equiv Z_{\L\L_0}\left[J,\bar{J};0\right]$.

$K_{\L\L_0} \to (1-K_{\L})$ for $\L_0 \to \infty$. The cut-off functions
$K_{\L}$ and $K_{\L\L_0}$ are chosen to be analytic functions; an
explicit representation for them will be given later.

The flow of the functional $Z_{\L\L_0}$ from $\L$ to zero can be
represented as
\begin{equation}\label{fL}
Z_{0\L_0}\left[J,\bar{J}\right]= 
\exp \left\{\int d^8 z \frac{\delta}{\delta \bar{J}}
\left[K_{\L_0}^{-1}-K_{\L\L_0}^{-1} - \chi K_{\L}K_{\L\L_0}^{-1}\right]
\frac{\delta}{\delta J} \right\}
Z_{\L\L_0}\left[J,\bar{J};\chi\right]
\end{equation}
Observe that $K_{\L\L_0}(p)$ goes to zero as $p^2/\L^2$ for 
$p^2/\L^2 \to 0$, so that 
the Wilsonian Green functions generated by
$Z_{\L\L_0}\left[J,\bar{J};0\right]$ are infrared finite for $\L > 0$
even at exceptional momenta.

The 1PI functional generator corresponding to 
$Z_{\L\L_0}=e^{W_{\L\L_0}}$ is obtained by Legendre transformation
\begin{equation}
\Gamma_{\L\L_0}\left[\Phi,\bar{\Phi};\chi\right] = -W_{\L\L_0}\left[J,\bar{J};\chi\right]+
\int d^6 z J \Phi + \int d^6 \bar{z} \bar{J} \bar{\Phi}
\end{equation}
with
\begin{equation}
\Phi = \frac{\delta W_{\L\L_0}}{\delta J}~~,~~
J = \frac{\delta \Gamma_{\L\L_0}}{\delta \Phi}
\end{equation}

Let us introduce a renormalization scheme in which some
renormalization scale $\L_R$ appears; according to general arguments
the Gell-Mann and Low renormalization group equation 
on $Z\left[J,\bar{J}\right] \equiv Z_{0\infty}\left[J,\bar{J}\right]$ holds
\begin{equation}\label{rge}
\left\{ \L_R \frac{\partial}{\partial \L_R} + 
\beta \frac{\partial}{\partial g}
+\gamma_{\Phi} \left[ \int d^6 z J\frac{\delta}{\delta J} +
\int d^6 \bar{z} \bar{J}\frac{\delta}{\delta \bar{J}}\right]\right\}
Z\left[J,\bar{J}\right] = 0
\end{equation}
where $\beta$ and $\gamma_{\Phi}$ are functions of $g$.
From eq. (\ref{fL}) it follows that 
$Z_{\L}\left[J,\bar{J};\chi\right]\equiv Z_{\L\infty}\left[J,\bar{J};\chi\right]=
e^{W_{\L}\left[J,\bar{J};\chi\right]}$ satisfies the effective 
renormalization group equation 
\begin{align}\label{Erge}
&\left\{ \L_R \frac{\partial}{\partial \L_R} + 
\beta \frac{\partial}{\partial g}
+\gamma_{\Phi} \left[ \int d^6 z J\frac{\delta}{\delta J} +
\int d^6 \bar{z} \bar{J}\frac{\delta}{\delta \bar{J}} \right] \right\}
Z_{\L}\left[J,\bar{J}\right] \nonumber \\
&=-2\frac{\partial}{\partial \chi}
Z_{\L}\left[J,\bar{J};\chi\right]_{\big|_{\chi=0}}
\end{align}
$W_{\L}\left[J,\bar{J};\chi\right]$ satisfies the same equation (\ref{Erge}) as
$Z_{\L}[J,\bar{J};\chi]$.
Making a Legendre transformation we get
\begin{equation}\label{Erg}
\left\{ \L_R \frac{\partial}{\partial \L_R} + 
\beta \frac{\partial}{\partial g}
-\gamma_{\Phi} \left[ \int d^6 z \Phi\frac{\delta}{\delta \Phi} +
\int d^6 \bar{z} \bar{\Phi}\frac{\delta}{\delta \bar{\Phi}}\right]\right\}
\Gamma_{\L}\left[\Phi,\bar{\Phi}\right] = 
2\gamma_{\Phi} {\cal T}_{\L}\left[\Phi,\bar{\Phi}\right]
\end{equation}
where
\begin{equation}\label{T}
{\cal T}_{\L}\left[\Phi,\bar{\Phi}\right]=
-\frac{\partial}{\partial \chi}
\Gamma_{\L}\left[\Phi,\bar{\Phi};\chi\right]_{\big|_{\chi=0}}
\end{equation}

Making a Volterra expansion 
$\Gamma_{\L\L_0}\left[\Phi,\bar{\Phi}\right]=
\sum_{n,\bar{n} \geq 0} \Gamma^{\L\L_0}_{n,\bar{n}}\left[\Phi,\bar{\Phi}\right]$,
where $\Gamma^{\L\L_0}_{n,\bar{n}}$ is the monomial of order $n ~(\bar{n})$
in $\Phi ~ (\bar{\Phi})$
and making a similar expansion on ${\cal T}_{\L}$ we get
\begin{equation}\label{ErgV}
\left\{ \L_R \frac{\partial}{\partial \L_R} + 
\beta \frac{\partial}{\partial g}
-(n+\bar{n})\gamma_{\Phi} \right\}
\Gamma^{\L}_{n,\bar{n}}\left[\Phi,\bar{\Phi}\right] = 
2\gamma_{\Phi} {\cal T}^{\L}_{n,\bar{n}}\left[\Phi,\bar{\Phi}\right]
\end{equation}
where $\Gamma_{\L} = \Gamma_{\L\infty}$.
The R-symmetry implies that $n-\bar{n} \equiv 0\, (mod. 3)$.
The only relevant vertices are 
$\Gamma_{1,1},\, \Gamma_{3,0},\,\Gamma_{0,3}$ .
In \cite{GSR} it was proven that all the 1PI graphs can be written
as integrals over a single $\int d^4\theta$; therefore
\begin{equation}\label{G2}
\Gamma^{\L\L_0}_{1,1}\left[\Phi,\bar{\Phi}\right] = \int d^4 \theta \int_p
\bar{\Phi}(\theta, -p)\Phi(\theta,p) F^{\L\L_0}(p^2)
\end{equation}
where $\int_p \equiv \int \frac{d^4p}{(2\pi)^4}$ and
$F^{\L\L_0(l)}$ is regular in $p=0$ for $\L > 0$.

Using the same theorem one can arrive at
\begin{equation}\label{G3}
\Gamma^{\L\L_0}_{3,0}\left[\Phi\right] = \int d^2 \theta \int_{p_1,p_2}
\Phi(\theta, p_1)\Phi(\theta,p_2) \Phi(\theta,-p_1-p_2)
G^{\L\L_0}(p_1,p_2)
\end{equation}
where $G^{\L\L_0(l)}(p_1,p_2)-c_2^{(l)}$ goes to zero for $p_1, ~p_2 \to 0$ as a
bilinear in $p_1$ and $p_2$, provided $\L > 0$.

This statement can be easily proven looking at each supergraph
contributing to $\Gamma^{\L\L_0(l)}_{3,0}$: apart from the
$l$-loop counterterm $c_2^{(l)}$, which is trivially of the form of
eq.(\ref{G3}),  due to the R-symmetry there are exactly two $D$'s more
than $\bar{D}$'s; using the manipulations on supergraphs explained in
\cite{GSR}, these two $D$'s are pulled out of the supergraph and 
act on the external legs, where they combine with
the two $\bar{D}$'s in $\int d^4\theta \simeq \int d^2\theta
\bar{D}^2$ to give two external momenta; since the remaining bosonic Feynman
integrals are clearly regular at zero momentum for $\L > 0$, the
assertion follows.

${\cal T}^{\L}_{1,1}$ and ${\cal T}^{\L}_{3,0}$ have the same
structure as $\Gamma^{\L}_{1,1}$ and $\Gamma^{\L}_{3,0}$
respectively;
\begin{equation}\label{T2}
{\cal T}^{\L}_{1,1}\left[\Phi,\bar{\Phi}\right] = \int d^4 \theta \int_p
\bar{\Phi}(\theta, -p)\Phi(\theta,p) T^{\L}(p^2)
\end{equation}
and ${\cal T}^{\L}_{3,0}$ gives a vertex which is vanishing at zero
momentum for $\L > 0$.

Imposing Wilsonian renormalization conditions at $\L=\L_R >0$ one can
use the effective renormalization group equation (\ref{Erg}) to give an
expression for the beta and gamma functions in terms of the Wilsonian
effective action. 

Due to the infrared finiteness of the Wilsonian Green functions at a
scale $\L_R > 0$, one can choose the following 
standard set of Wilsonian renormalization conditions (HS scheme):
\begin{align}\label{rc}
F^{\L_R\L_0}(0)=-1&& G^{\L_R\L_0}(0,0)= \frac{g}{3!}
\end{align}
which imply the non-renormalization of the chiral superpotential:
$c_2^{(l)}= G^{\L_R\L_0(l)}(0,0) = 0$ for all $l \geq 1$.

Using eqs.(\ref{ErgV}-\ref{rc}) we get
\begin{equation}\label{gamma}
\gamma_{\Phi} =\frac{1}{2\left[1-T^{\L_R}(0)\right]} \L \frac{\partial}{\partial
\L} {F^{\L}(0)}_{\big|_{\L_R}}
\end{equation}
and
\begin{equation}\label{beta}
\beta = 3g \gamma_{\Phi}
\end{equation}
\begin{figure}
\begin{center}
\includegraphics[width=0.95\textwidth]{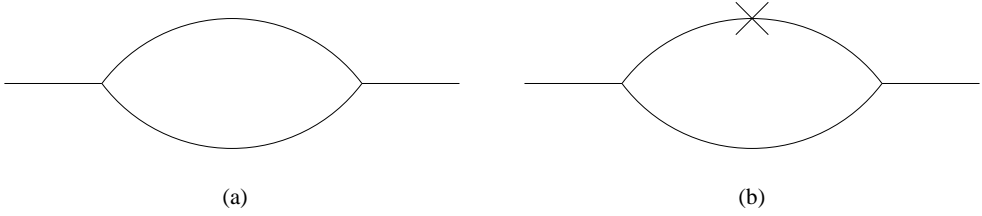}
\end{center}
\caption{one-loop contributions}  
\label{oneloop}
\end{figure}
According to general arguments one expects that the beta function is
scheme-independent at the first two loops. Using eq.(\ref{beta}) one can
reduce the computation of beta to the computation of gamma.
Let us compute $\gamma_{\Phi}$ at the first two loops.
At one loop the only non-vanishing bare parameter is
\begin{equation}
c_1^{(1)} = \frac{g^2}{2} \int_q D_{\L_R\L_0}^2(q)
\end{equation}
where $D_{\L_R\L_0}(q)\equiv K_{\L_R\L_0}(q)/q^2$ is the hard
scalar propagator; this counterterm cancels the self-energy graph (see
Fig.\ref{oneloop}a) at scale $\L_R$ and at zero momentum.

One has
\begin{equation}\label{F1}
F^{\L\L_0(1)}(0) = -\frac{g^2}{2} \int_q 
\left[D^2_{\L\L_0}- D^2_{\L_R\L_0}\right](q)
\end{equation}

At one loop not only $c_2^{(1)}=G^{\L_R\L_0(1)}(0,0)=0$, but also
one has at arbitrary momentum $G^{\L_R\L_0(1)}(p_1,p_2)=0$ .

Finally we get (with $T^{\L} \equiv T^{\L\infty}$)
\begin{equation}
T^{\L_R\L_0(1)}(0) =-g^2  \int_q K_{\L_R}(q) D^2_{\L_R\L_0}(q)
\end{equation}
corresponding to the Wilsonian graph in Fig.1b , where the cross
indicates the insertion of the $\chi$-term of eq.(\ref{Lbare}).

Let us consider the class of HS schemes characterized by a cut-off of
the form
\begin{equation}\label{K}
K_{\L\L_0 }(p) = p^2 \int_{\L_0^{-2}}^{\infty} d \alpha\  e^{-\alpha
p^2} \rho (\alpha \L^2)
\end{equation}
where the function $\rho(x)$ satisfies $\rho(0)=1$ and goes to zero
fast enough for $x \to \infty$. 
In the present case it is not necessary to add the
condition $\rho'(0)=0$ on the cut-off function (\ref{K}), required in \cite{PRR,PR},
since in the Wess-Zumino model the quadratic divergences cancel.

From eq.(\ref{gamma},\ref{F1}) we get
\begin{align}
& \gamma_{\Phi}^{(1)} = \frac{1}{2}\L \frac{\partial}{\partial
\L} F^{\L(1)}(0)_{\big|_{\L_R}}=  \\
& \lim_{\L_0 \to \infty}
\frac{-g^2}{16\pi^2} \int_{\frac{\L^2}{\L_0^2}}^{\infty}
d\alpha_1d\alpha_2\frac{\alpha_1}{(\alpha_1+\alpha_2)^2}
\rho'(\alpha_1)\rho(\alpha_2) = \frac{1}{2} \frac{g^2}{16\pi^2} \nonumber
\end{align}
\begin{figure}
\begin{center}

\includegraphics[width=0.45\textwidth]{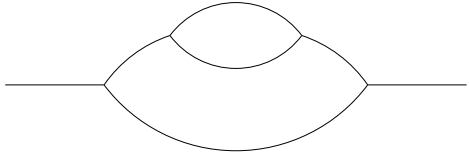}
\caption{two-loops contribution}
\label{twoloops}
\end{center}
\end{figure}
At two loops one has (see Fig.\ref{twoloops})
\begin{align}\label{F2}
&F^{\L\L_0(2)}(0) =\\& \frac{g^4}{2} \int_{pq} K_{\L\L_0}(p)
D^2_{\L\L_0}(p)\left[ D_{\L\L_0}(q)D_{\L\L_0}(p+q)-D_{\L_R\L_0}^2(q)\right]
+ c_1^{(2)}\nonumber
\end{align}

Observing that
\begin{equation}
\L {\frac{\partial}{\partial\L}}_{\big|_{\L_R}}
\int_{pq} \left(1-K_{\L\L_0}(p)\right)
D^2_{\L\L_0}(p)\left[ D_{\L\L_0}(q)D_{\L\L_0}(p+q)-D_{\L\L_0}^2(q)\right]
\end{equation}
vanishes in the limit $\L_0 \to \infty$
we find, using eqs.(\ref{gamma},\ref{F2}),
\begin{align}
&\gamma_{\Phi}^{(2)} =\frac{1}{2} \lim_{\L_0 \to \infty} 
\L {\frac{\partial}{\partial\L}}_{\big|_{\L_R}}
\left[F^{\L\L_0(2)}(0)+T^{\L_R(1)}(0)F^{\L\L_0(1)}(0)\right]= \\
& \frac{g^4}{4} \L {\frac{\partial}{\partial\L}}_{\big|_{\L_R}}
\int_{pq} 
D^2_{\L\L_0}(p)\left[ D_{\L\L_0}(q)D_{\L\L_0}(p+q)-D_{\L_R\L_0}^2(q)\right]
= -\frac{1}{2}\left(\frac{g^2}{16\pi^2}\right)^2 \nonumber
\end{align}
where in the last step the integral is the same as for computing
$\beta^{(2)}$ in $g\phi^4$ in the HS scheme \cite{PR} so that we get
the expression of the beta function at the first two loops
\begin{equation}
\beta =3g\left[\frac{1}{2}\frac{g^2}{16\pi^2}-
\frac{1}{2}\left(\frac{g^2}{16\pi^2}\right)^2\right]
\end{equation}
in agreement with \cite{AG}.

\begin{figure}
\begin{center}
\includegraphics[width=0.80\textwidth]{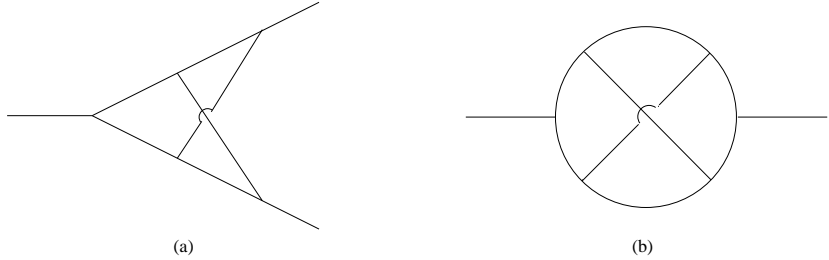}
\caption{three-loops contributions}
\label{threeloops}
\end{center}
\end{figure}

Let us compare the HS renormalization scheme with a renormalization scheme
at $\L=0$ with non-zero momentum subtraction points
\begin{align}
F^{\L=0 \L_0}(\mu^2)=-1&&
G^{\L=0 \L_0}(\bar{p}_1,\bar{p}_2)= \frac{g}{3!}
\end{align}
For generic non-zero momenta $G^{\L=0 \L_0(l)}(p_1,p_2)$ is
non-vanishing for $l \geq 2$; e.g. $G^{\L=0 \L_0(l)}(0,p)$ has been
evaluated in \cite{JJW}, corresponding to the graph in Fig.\ref{threeloops}a; this
chiral vertex gives also a non-vanishing contribution to the
three-loop self-energy graph of Fig.\ref{threeloops}b, which has been evaluated in
\cite{AG,Sen}. This shows that $G^{\L=0 \L_0(l)}(s,p)$ is
non-vanishing not only at $s=0$, but also for small $s$. 
Observe furthermore that the finiteness of $G^{\L=0 \L_0(l)}(0,p)$ for
$p \to 0$ and 
$l=2$ is accidental. In fact a simple renormalization group argument
shows that at three loops it diverges as $\ln\left( p / \L_R\right)$ for $p \to 0$,
due to the diagram of Fig.\ref{threeloops}a with a self-energy insertion. 

Using the notation of \cite{AG},
the bare coupling constant has the form
\begin{equation}
g_0 = \mu^{\epsilon} f(g) Z_{\Phi}^{-3/2}
\end{equation}
where
$\mu^{\epsilon}$ is the usual factor introduced in dimensional regularization
and $f(g)=g+...$ is an odd function of $g$, which is finite due to the
non-renormalization theorem (in its `weak' form assuring that the
chiral supergraphs are superficially convergent) but not equal to $g$,
due to the above-mentioned quantum corrections to the superpotential.
One gets
\begin{equation}
\gamma_{\Phi}=\frac{1}{2}{\mu \frac{\partial}{\partial
\mu}}_{\big|g_0}\ln Z_{\Phi}
\end{equation}
\begin{equation}
\beta = \mu{\frac{\partial}{\partial \mu}}_{\big|g_0}g=
3\gamma_{\Phi}\left[{\frac{d}{d g}\ln f}\right]^{-1}
\end{equation}
The beta and gamma functions are still proportional, but their
relation involves the function $[\ln f]'(g)$ which has to be computed
order by order in perturbation theory in a generic
momentum-subtraction scheme, so that in general 
$\beta^{(l)} \neq 3 g \gamma^{(l)}$ for $l \geq 3$.
In fact the same is true in the massive Wess-Zumino model
\cite{Piguet}; only choosing renormalization conditions at zero
momentum one gets the relation (\ref{beta}) in a natural way
(i.e. without order-by-order fine tuning of the renormalization
conditions).

In \cite{AG} a `momentum subtraction' scheme is also considered, in which
only the renormalization condition $F^{\L=0}(\bar{p}^2)=-1$ is imposed,
while the vertex is not renormalized, relying on a supersymmetric
invariant regularization and imposing the condition 
$g_0 =\mu^{\epsilon} gZ^{-3/2}$;
then the relation (\ref{beta}) follows. This procedure is consistent due to
the non-renormalization theorem, but it relies on the choice of an
invariant regulator.

The advantage of a renormalization scheme in which all the relevant
vertices are subjected to a renormalization condition is that the
renormalized theory is completely defined, regardless of the choice of
an invariant regulator.
Using the HS scheme (with $\rho'(0)=0$)
with a generic (non supersymmetric invariant) ultraviolet
cut-off, imposing that the renormalization conditions (\ref{rc}) are
satisfied in the limit of infinite ultraviolet cut-off, one obtains
the same renormalized Green functions as in the case previously
studied, and hence the relation (\ref{beta}).

Using the Wilson-Polchinski \cite{Wil,Polchi}
flow equation technique for the
Wilsonian effectice functional, simple rigorous proofs of
renormalizability and other important results in perturbation theory
have been obtained \cite{flow}. It would be interesting to use this
approach to study the massless Wess-Zumino model. A first step in this
direction has been made in \cite{BV}, where the flow equation has been
written in the superfield formalism.

Let us make a comment on the question of the `holomorphic anomaly'.
In this letter we chose the coupling constant $g$ to be real; taking
it complex the conclusions are similar; choosing Wilsonian
renormalization conditions at zero momentum the `holomorphic anomaly'
term $\int d^6z g^3\bar{g}^2\Phi^3$ appearing at two loops,
corresponding to the contribution of Fig. 3a \cite{JJW} is avoided as
long as the Wilsonian scale $\L$ is different from zero. 
This is in agreement with
\cite{Sei}, where it was observed that using a Wilsonian effective action
these `anomalies' are avoided, as first suggested in \cite{Shif} in
the context of supersymmetric gauge theories.

\end{document}